\documentclass[sigconf, screen, authorversion]{acmart}

\AtBeginDocument{%
  }

\copyrightyear{2025}
\acmYear{2025}
\setcopyright{cc}
\setcctype{by-nc-sa}
\acmConference[ITiCSE 2025]{Proceedings of the 30th ACM Conference on
Innovation and Technology in Computer Science Education V. 1}{June 27-July 2,
2025}{Nijmegen, Netherlands}
\acmBooktitle{Proceedings of the 30th ACM Conference on Innovation and
Technology in Computer Science Education V. 1 (ITiCSE 2025), June 27-July 2,
2025, Nijmegen, Netherlands}\acmDOI{10.1145/3724363.3729094}
\acmISBN{979-8-4007-1567-9/2025/06}

\usepackage[utf8]{inputenc}
\usepackage{colortbl}
\usepackage{subcaption}
\usepackage{diagbox}
\usepackage{multirow}
\usepackage{xspace}
\usepackage{enumitem}
\usepackage{mdframed}
\usepackage{xcolor} 

\usepackage{caption}
\usepackage{subcaption}
\usepackage{wrapfig}
\usepackage{etoolbox}
\usepackage{verbatim}
\usepackage{amsmath}
\usepackage{graphicx}
\usepackage{tabularray}
\usepackage{tcolorbox}
\usepackage{listings}

\begin{document}

\title[Prompt Programming]{Prompt Programming: A Platform for Dialogue-based Computational Problem Solving with Generative AI Models}

\author{Victor-Alexandru P{\u a}durean}
\affiliation{
  \institution{MPI-SWS}
  \city{Saarbr{\"u}cken}
  \country{Germany}  
}
\email{vpadurea@mpi-sws.org}

\author{Paul Denny}
\affiliation{
  \institution{University of Auckland}
  \city{Auckland}
  \country{New Zealand}  
}
\email{paul@cs.auckland.ac.nz}

\author{Alkis Gotovos}
\affiliation{
  \institution{MPI-SWS}
  \city{Saarbr{\"u}cken}
  \country{Germany}  
}
\email{agkotovo@mpi-sws.org}

\author{Adish Singla}
\affiliation{
  \institution{MPI-SWS}
  \city{Saarbr{\"u}cken}
  \country{Germany}  
}
\email{adishs@mpi-sws.org}


\begin{abstract}
\looseness-1Computing students increasingly rely on generative AI tools for programming assistance, often without formal instruction or guidance. This highlights a need to teach students how to effectively interact with AI models, particularly through natural language prompts, to generate and critically evaluate code for solving computational tasks. To address this, we developed a novel platform for prompt programming that enables authentic dialogue-based interactions, supports problems involving multiple interdependent functions, and offers on-request execution of generated code. Data analysis from over $900$ students in an introductory programming course revealed high engagement, with the majority of prompts occurring within multi-turn dialogues. Problems with multiple interdependent functions encouraged iterative refinement, with progression graphs highlighting several common strategies. Students were highly selective about the code they chose to test, suggesting that on-request execution of generated code promoted critical thinking. Given the growing importance of learning dialogue-based programming with AI, we provide this tool as a publicly accessible resource, accompanied by a corpus of programming problems for educational use.
\end{abstract}



\maketitle

\section{Introduction}\label{sec.introduction}

\looseness-1Generative AI has transformed programming and the skills required to excel in this field \cite{denny2024cacm}. Students are increasingly turning to AI tools to assist with programming tasks, often without any formal guidance \cite{DBLP:conf/kolicalling/AmoozadehNPAPHR24}. This highlights a need for educators to actively teach the new skills required to use these tools effectively. Traditionally, introductory courses focused on teaching students to write code by transforming clear problem statements into syntactically and semantically correct code. However, with large language models (LLMs) now capable of solving this step, there is a growing need to emphasize reading, critically evaluating generated code, and crafting precise natural language prompts to achieve desired outcomes.

\looseness-1Several tools were recently described to help students develop prompt-crafting skills. For instance, CodeAid expands prompts into high-level pseudocode to help students structure solutions while programming \cite{kazemitabaar2024codeaid}, whereas an EiPE-inspired tool converts prompts into code fragments to generate feedback for code explanation tasks \cite{smith2024prompting}. A more direct approach is the concept of `Prompt Problems' which challenge students to create single-shot prompts to solve computational tasks presented visually \cite{denny2024prompt}. Building on this idea, in the current work we propose several new features that improve the value of this learning-by-prompting approach: dialogue-based interactions, support for problems involving multiple functions, and on-request execution of generated code. These novel features support more authentic interactions with LLMs, promote critical evaluation of code, and enable iterative, real-world problem-solving.

\begin{figure*}[t!]
    \centering
    \begin{subfigure}{0.405\textwidth}
        \centering
        \begin{tcolorbox}[colback=gray!10, colframe=gray!70!black, boxsep=0mm]
            \small\looseness-1
            \input{figs/illustration_multi/description.tex}
        \end{tcolorbox}
        \includegraphics[width=\textwidth,trim={0mm 40mm 0mm 3mm},clip]{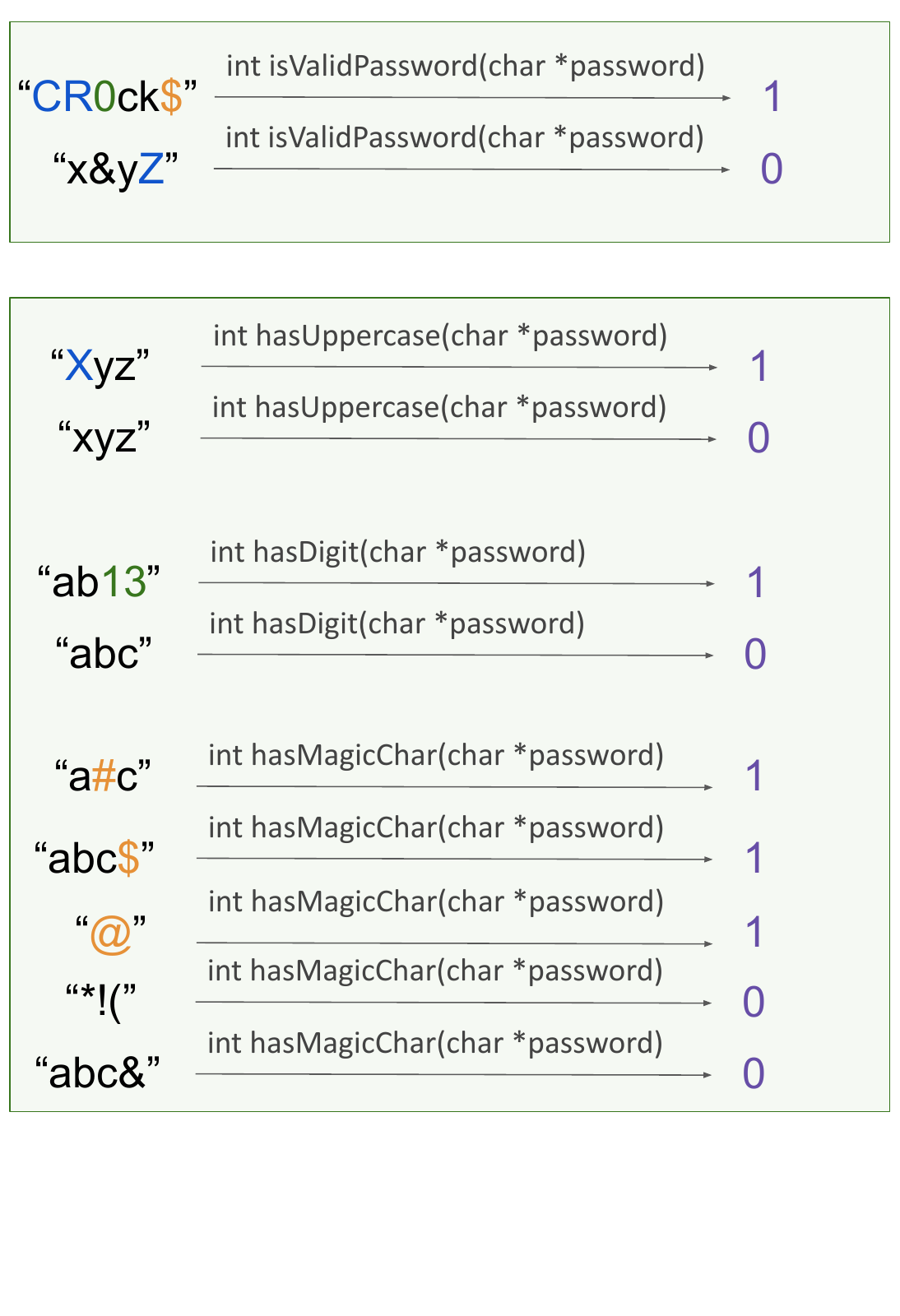}
        \caption{Problem description and input-output specifications.}
        \vspace{-2mm}
        \label{fig.illustration_multi.description}
    \end{subfigure}
    \hfill
    \begin{subfigure}{0.515\textwidth}
        \centering
        \includegraphics[width=\textwidth]{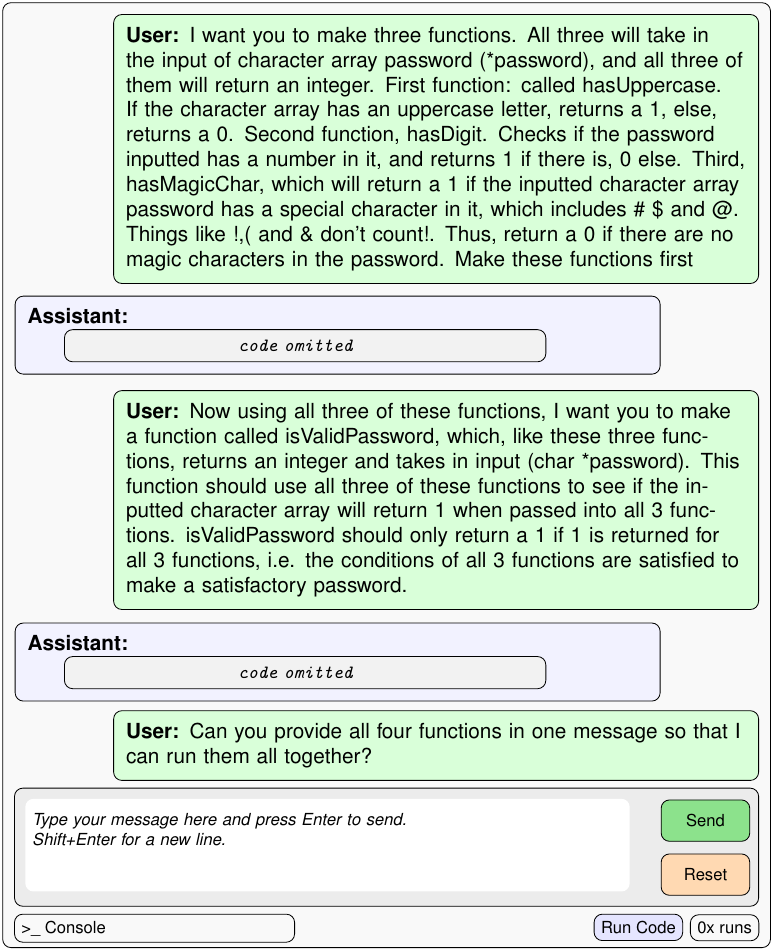}
        \caption{Example of a student's interaction through the chat interface.}
        \vspace{-2mm}
        \label{fig.illustration_multi.interaction}
    \end{subfigure}
    \caption{\looseness-1Illustration for a \emph{multi-function} problem. (a) shows the problem description, including a visual depiction of multiple function specifications for the `password validation' problem. (b) recreates the chat interface, showing a genuine student's successful attempt at interacting with the LLM to solve the problem (generated code and last assistant message omitted).}
    \label{fig.illustration_multi}
\end{figure*}

In this paper, we introduce \emph{Prompt Programming}, a free web-based platform for students to practice crafting and refining prompts for LLMs, available at \url{https://www.promptprogram.org} with support for Python and C-based exercises. Unlike previous single-shot approaches, it enables multi-turn conversations for iterative refinement, provides on-request execution to test code (Figure~\ref{fig.illustration_single}), and supports multi-function problems (Figure~\ref{fig.illustration_multi}). We evaluate these novel features using student data from a large introductory programming course, guided by three research questions:


\begin{figure*}[t!]
    \centering
    \begin{subfigure}{0.405\textwidth}
        \centering
        \begin{tcolorbox}[colback=gray!10, colframe=gray!70!black, boxsep=0mm]
            \small
            \input{figs/illustration_single/description.tex}
        \end{tcolorbox}
        \includegraphics[width=0.96\textwidth,trim={2mm 51mm 2mm 5mm},clip]{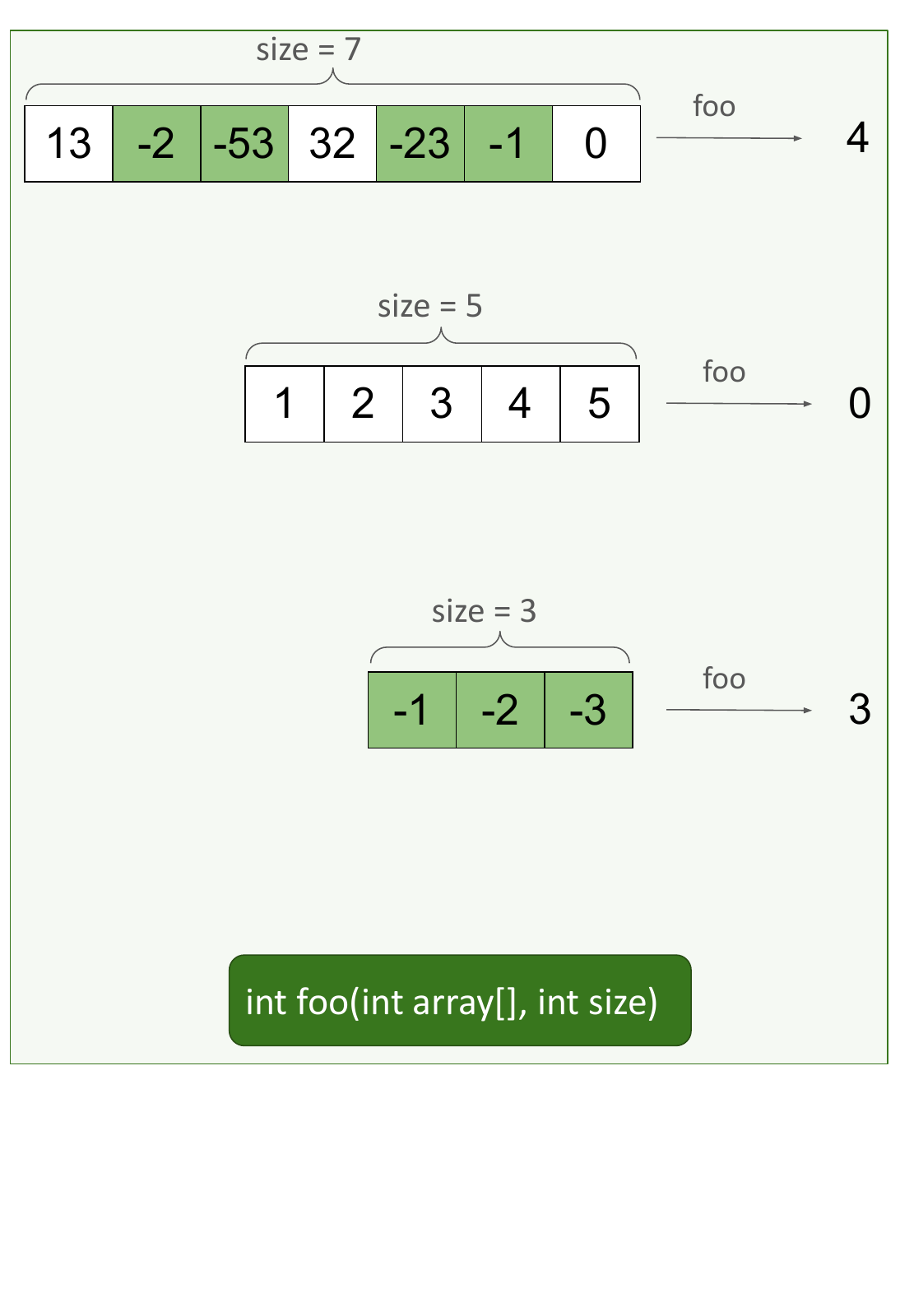}
        \caption{Problem description and input-output specifications.}
        \vspace{-2.5mm}
        \label{fig.illustration_single.description}
    \end{subfigure}
    \hfill
    \begin{subfigure}{0.52\textwidth}
        \centering
        \includegraphics[width=\textwidth]{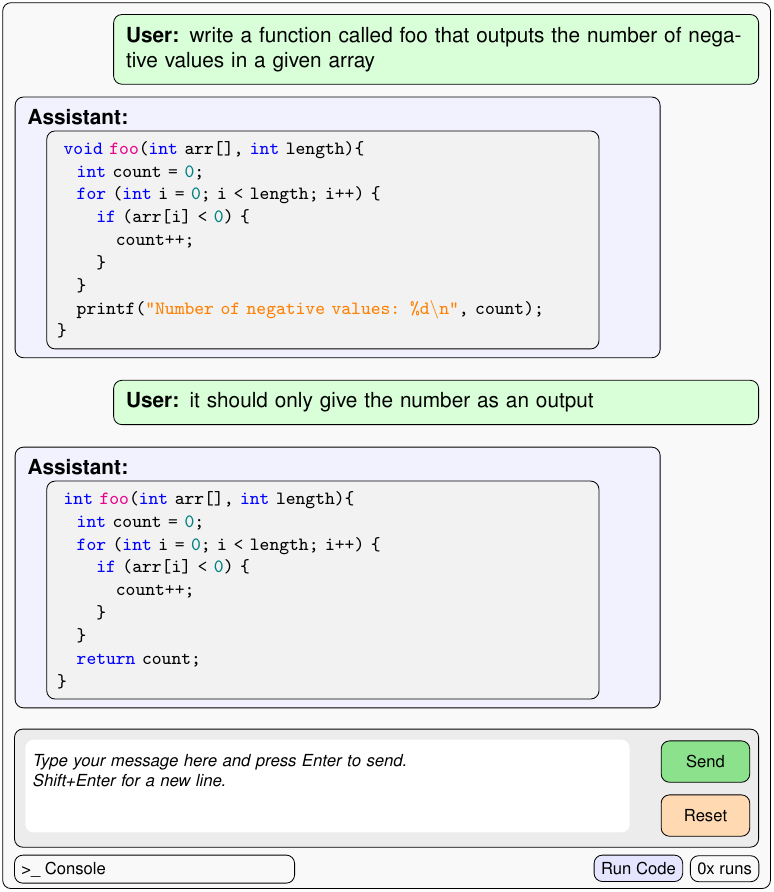}
        \caption{Example of a student's interaction through the chat interface.}
        \vspace{-2.5mm}
        \label{fig.illustration_single.interaction}
    \end{subfigure}
    \caption{Illustration for a \emph{single-function} problem. (a) shows the problem description, including a visual depiction of the input-output specifications for the `count negatives in an array' problem. (b) recreates the chat interface, showing a genuine student's successful attempt at interacting with the LLM to solve the problem via multiple messages.}
    \label{fig.illustration_single}
\end{figure*}


\setlength{\leftmargini}{1em}
\begin{itemize} 
\item RQ1: What approaches do students employ when solving multi-function problems?
\item RQ2: How does problem complexity influence the length and message size of multi-turn conversations?
\item RQ3: When using on-request execution, how selective are students in deciding whether to execute generated code? 
\end{itemize}
\section{Related work}\label{sec.relatedwork}
\looseness-1\textbf{Generative AI for computing education.} Recent advances in generative AI, in particular, code-generating large language models, have the potential to drastically change the landscape of computing and programming education~\cite{denny2024cacm,prather2023navigating}. Recent work has demonstrated this potential in numerous programming education scenarios, including repairing buggy programs~\cite{DBLP:journals/pacmpl/ZhangCGLPSV24}, enhancing programming-error-messages~\cite{DBLP:conf/sigcse/WangMP24,DBLP:conf/sigcse/0001HSRDPB23,DBLP:conf/edm/PhungCGKMSS23}, providing code explanations~\cite{DBLP:conf/icer/SarsaDH022,macneil23sigcse}, and generating new programming exercises~\cite{DBLP:conf/icer/SarsaDH022,padurean2024neural} or questions~\cite{doughty2024comparative,li2023leveraging}. Moreover, generative AI has been leveraged to create debugging exercises that help learners identify and fix code errors~\cite{ma2024hypocompass,padurean2024bugspotter} or for enabling conversational programming~\cite{denny2023conversing}. It can also act as an AI pair programmer, providing real-time feedback~\cite{lohr2025youre,zamfirescupereira2024conversational} and fostering collaboration~\cite{DBLP:conf/chi/MozannarBFH24,DBLP:conf/aied/MaWK23,DBLP:conf/chi/Vaithilingam0G22,DBLP:conf/iticse/LiuYHBBL24}, or support mixed-modality tasks with both text and code~\cite{rawal2024hints,ahmed2025feasibility}. These advancements highlight the potential of generative AI to address diverse educational challenges and offer unique support in novice programming environments.

\textbf{Prompt problems.} Denny et al.~\cite{denny2024prompt} introduced the idea of `Prompt Problems', a novel exercise where students craft natural language prompts to solve computational tasks presented visually. Accompanied by a prototype tool called \emph{Promptly}, this approach evaluates code generated from student-written prompts against a test suite. Implemented in CS1 and CS2 courses, their study revealed positive student engagement, with learners appreciating how these exercises foster computational thinking and expose them to new constructs. However, the prototype tool presented in this prior work supports only single-turn prompts, which limits how well it can be applied to more complex, real-world tasks. 

Kerslake et al.~\cite{kerslake2024integrating} investigated the integration of natural language prompting tasks into introductory programming courses, shifting the focus from syntax mastery to problem-solving. In addition to Prompt Problems, another type of task was introduced that focused on code explanation -- students crafted prompts to generate code equivalent to provided code fragments. Their study found that these tasks engaged a broader range of cognitive skills compared to traditional programming assessments and appealed to students who found syntax challenging. The work highlights the potential of prompt-focused activities to reduce barriers for novices and encourage diverse participation in programming.

Most recently, Prather et al.~\cite{prather2024breaking} explored the use of Prompt Problems in multilingual settings to address barriers faced by non-native English-speaking (NNES) students in programming education. The study found that NNES students successfully used their native languages for solving computational tasks through LLMs, although challenges in terminology persisted. Moreover, this work utilized the same prototype tool described by Denny et al.~\cite{denny2024prompt}, which was limited to single-function problems and single-prompt interactions, and is not publicly available.

\textbf{Generative AI-based conversational agents for education.}
Generative AI conversational agents are increasingly recognized for their transformative potential across diverse educational contexts~\cite{DBLP:journals/corr/abs-2402-01580}. These agents are used to simulate classroom environments for teacher training~\cite{DBLP:conf/lats/MarkelOLP23,lee2024generative}, enable students to enhance their understanding by teaching AI peers~\cite{DBLP:conf/aied/SchmuckerXAM24}, and model student behavior~\cite{DBLP:conf/edm/NguyenTS24}. They can help in boosting creativity~\cite{acun2024gaienhanced,bailis2024wordplay} and critical thinking~\cite{DBLP:conf/lak/SonkarCLLMB24} through interactive tasks, puzzles, and problem-solving, demonstrating their versatility in reshaping education.

\section{Methodology}\label{sec.methodology}

This section describes our novel \emph{Prompt Programming} platform.
We highlight its core functionality, then detail our classroom deployment and our approach to analyzing student interactions.

\subsection{Prompt Programming Platform}

\begin{figure*}[t!]
    \centering
    \begin{subfigure}[t]{0.49\textwidth}
        \centering
        \includegraphics[width=\textwidth,trim={10mm 4mm 5.5mm 8mm},clip]{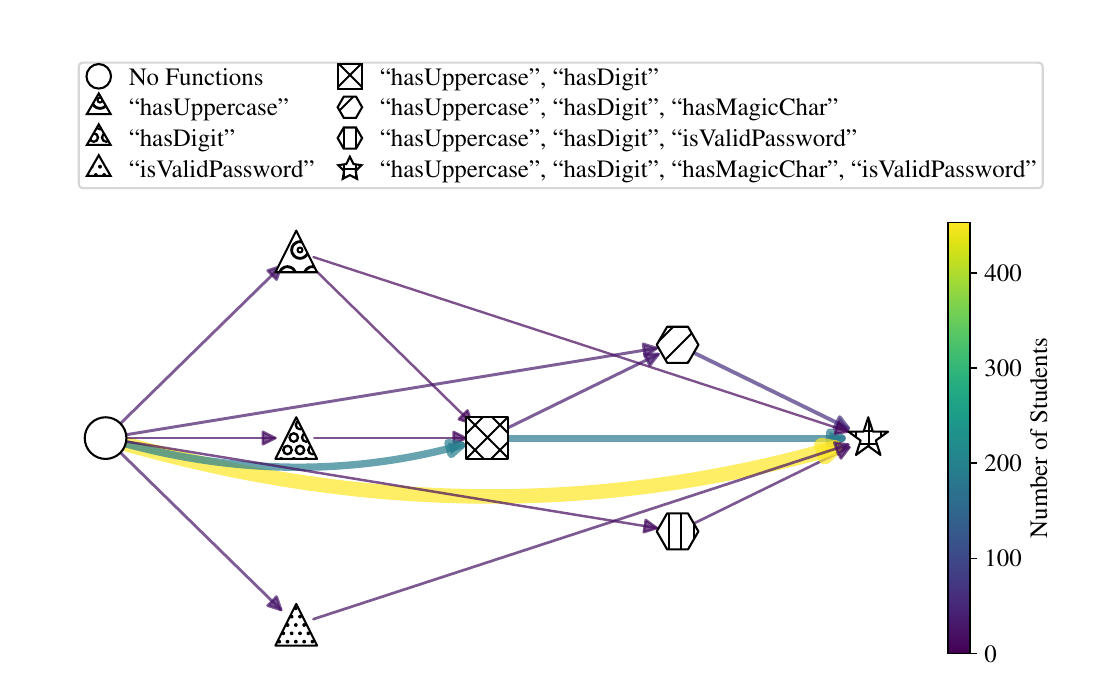}
        \caption{Progression for password validation problem solved by $781$ students.}
        \label{fig.results-multifunc.password}
    \end{subfigure}
    \hfill
    \begin{subfigure}[t]{0.49\textwidth}
        \centering
        \includegraphics[width=\textwidth,trim={10mm 4mm 5.5mm 8mm},clip]{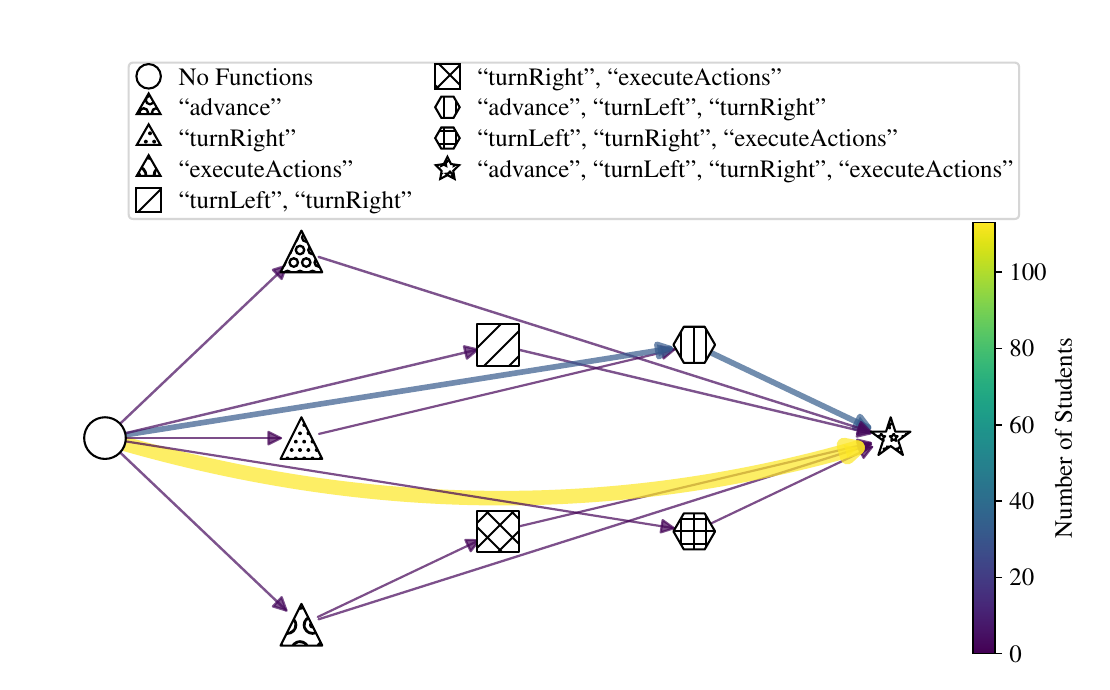}
        \caption{Progression for robot navigation problem solved by $156$ students.}
        \label{fig.results-multifunc.robot}
    \end{subfigure}
    \vspace{-2mm}
    \caption{
    Progression graphs for solving multi-function problems, highlighting the most common paths students followed when solving two of the \emph{multi-function} problems. Nodes represent sets of correctly implemented functions, and edges indicate transitions, with color gradients representing the number of students following each path.
    }
    \label{fig.results-multifunc}
\end{figure*}

\looseness-1We developed the \emph{Prompt Programming} platform to support iterative problem-solving with LLMs, focusing on three core functionalities: multi-function problems, multi-turn conversations, and on-request code execution. These allow students to iteratively refine prompts, critically evaluate model outputs before code execution, and develop a systematic approach to solving complex programming problems. By promoting iterative refinement of prompts and deeper engagement with code, the platform fosters critical thinking and problem-solving skills. We detail each functionality below.

First, the platform introduces \textbf{multi-function problems}, which allows students to practice solving complex tasks, mirroring real-world usage of generative models. To help students gradually adapt, we start with \emph{single-function} problems, inspired by earlier works \cite{denny2024prompt}, as illustrated in Figure~\ref{fig.illustration_single.description}. Building on this foundation, we introduce \emph{multi-function} problems, that challenge students to guide the LLM in implementing interdependent functions and creating structures or classes. Input-output specifications are provided for each function, as illustrated in Figure~\ref{fig.illustration_multi.description}, so students understand expected behaviors. A problem is considered solved only when all functions and constructs meet their respective specifications, encouraging students to ensure all components work together cohesively.

\looseness-1Second, the platform provides a chat-like interface for natural, \textbf{multi-turn conversations}, enabling students to iteratively refine their prompts, as shown in Figures~\ref{fig.illustration_single.interaction} and~\ref{fig.illustration_multi.interaction}. Each message is processed  with the full conversation history to generate context-aware responses. Besides the student's verbatim message, we prepend a system prompt specifying the programming language and instructing the LLM to exclude test code or a main function. A `Reset' button lets students clear the conversation history and start afresh.

\looseness-1Finally, the \emph{Prompt Programming} platform includes \textbf{on-request code execution}, enabling students to critically evaluate generated code before running it against hidden test cases. With the `Run Code' button, students can execute the most recently generated block of code. A counter tracks code execution requests, incentivizing students to be selective in using this feature. This functionality fosters code comprehension and critical judgment by encouraging students to hypothesize code behavior and address issues before running it.

\begin{figure*}[t!]
    \centering
    \begin{subfigure}[t]{0.49\linewidth}
        \centering
        \includegraphics[width=\linewidth]{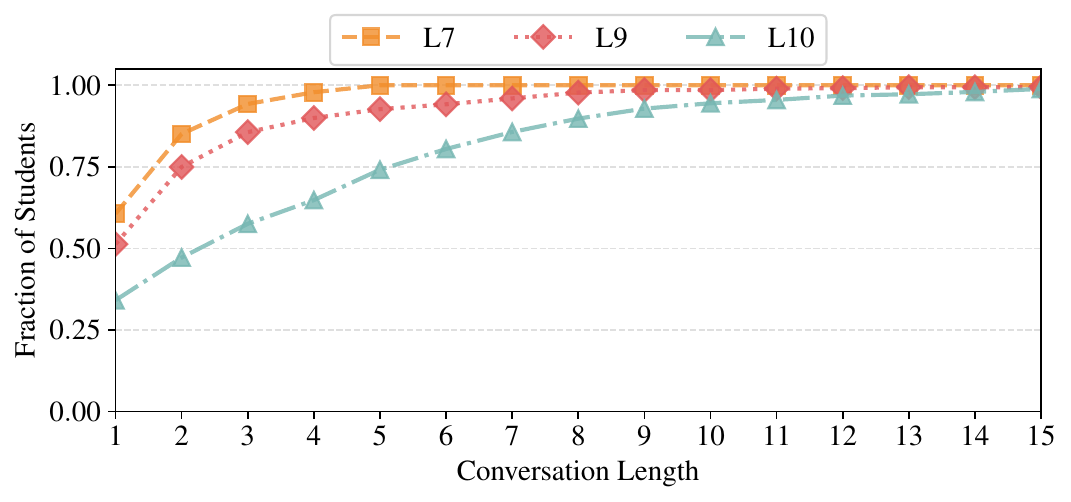}
        \caption{Fraction of students bucketed by successful conversation length.}
        \label{fig.results_multiturn.success}
    \end{subfigure}
    \hfill
    \begin{subfigure}[t]{0.49\linewidth}
        \centering
        \includegraphics[width=\linewidth]{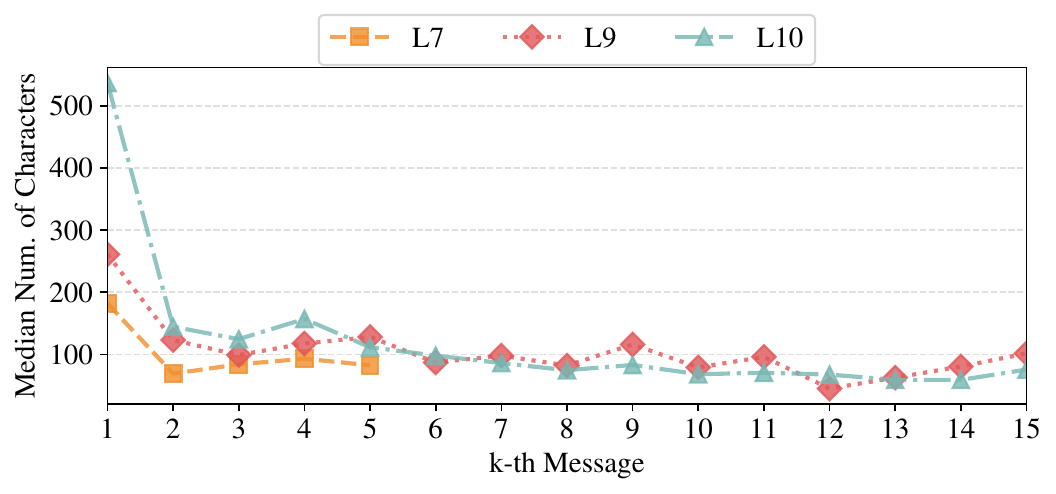}
        \caption{Median message size for successful conversations.}
        \label{fig.results_multiturn.effort}
    \end{subfigure}
    \vspace{-3mm}
    \caption{Usage of multi-turn conversations in solving problems across laboratory assignments. (a) shows the fraction of students achieving success within a given conversation length in their first successful conversation. (b) shows the median character length of messages by position (k-th message) during the first successful conversation.
    }
    \label{fig.results_multiturn}
\end{figure*}

\subsection{Study Setup}

In order to investigate how our platform supports students in prompt writing and iterative problem solving, we deployed a version of our \emph{Prompt Programming} platform, specifically focusing on C exercises, in a large introductory C programming course taught at the University of Auckland with $1,031$ students. The platform used GPT-4o mini \cite{GPT4omini}, popular for its accessibility and low cost. The primary goal of the deployment was to collect data on how students used features like multi-turn conversations and on-request code execution to tackle progressively complex problems, including multi-function problems that mirror real-world programming scenarios. To evaluate their engagement and usage patterns, we collected interaction data, including student queries, the LLM's responses, and code execution requests. Students were also invited to provide optional open-response feedback, providing qualitative insights into their interactions with \emph{Prompt Programming}.

\looseness-1The study spanned three laboratory assignments (L7, L9, and L10), each lasting one week and each comprising $3$ problems created by the authors, who have extensive teaching experience in programming. The problems were chosen to align with the topics students were learning during the course at the time of each laboratory. In L7, students needed to solve all $3$ simple \emph{single-function} problems: counting negatives in an array (see Figure~\ref{fig.illustration_single}), summing even numbers in an array, and finding the index of the last zero in an array. In L9, students needed to solve at least $1$ of the $3$ more complex \emph{single-function} problems: sorting a subarray within specified indices, updating a matrix by propagating $1$s across rows and columns, and adding binary numbers represented as arrays. Finally, in L10, students needed to solve at least $1$ of the $3$ \emph{multi-function} problems, which required helper functions and structures: password validation (see Figure~\ref{fig.illustration_multi}), guiding a robot's movement on a grid using a structure and functions for turning and advancing, and checking if an expression's round parentheses are balanced using a stack and corresponding functions. To prevent students from getting sidetracked within a single conversation, we limited conversations to $5$ messages for L7, and $20$ for L9 and L10 (conversations would need to be reset by the student once reaching these limits).


\section{Results}\label{sec.results}

Next, we summarize participation, present results for each research question (RQ), and highlight key insights from student reflections.

\subsection{Descriptive Statistics}

\looseness-1We begin by reporting the success rate of students across three labs. Recall that L7 required solving $3$ out of $3$ problems, while L9 and L10 required solving only $1$ out of $3$ problems. Out of $1,031$ students in the course, typically about $90\%$ of students participate in a given laboratory assignment. For students who participated in a given lab, the success rates are as follows: $96\%$ for L7, $96\%$ for L9, and $97\%$ for L10.  The average number of conversations and average total messages by a student for a problem in a given lab are: (i) $4$ and $5$ for L7, (ii) $4$ and $10$ for L9, and (iii) $2$ and $12$ for L10.
Interestingly, the success rates are high even for difficult problems, though the average number of messages increased with lab difficulty.

\subsection{RQ1: Solving Multi-function Problems}

\looseness-1To understand how students approach multi-function problems, we analyzed the progression of correctly implemented functions in conversations from L10. We took the first successful conversation for each student (i.e., the first conversation where they managed to solve the problem) and modeled progression as a directed graph, where nodes represent cumulative states of correctly implemented functions. A transition was recorded as an edge whenever new functions were added to the set of correctly implemented functions. We calculated edge weights based on the number of students whose conversations included a specific transition, highlighting the most frequent paths. To simplify visualization, we filtered the graph to include the top $15$ edges in terms of weights, focusing on the most common transitions students followed to solve multi-function problems.

Figures~\ref{fig.results-multifunc.password} and~\ref{fig.results-multifunc.robot} show that many students attempted to implement all functions in one step. However, the figures also highlight a large diversity of paths, with students using multi-turn conversations to iteratively refine functions and address errors while retaining correct components. Both problems revealed a hub-like intermediate state, where transitions were concentrated. This state often marked a point where most helper functions were correctly implemented, providing a foundation for tackling advanced aspects of the problem. These findings highlight that multi-function problems do encourage diverse problem-solving strategies -- in Section~\ref{sec:limitations} we discuss how support for the stepwise development of solutions could be better enabled in future work. 

\begin{tcolorbox}[colback=gray!10, colframe=black!40, rounded corners, boxsep=0mm, left=0.7mm, right=0.7mm, top=0.5mm,bottom=0.5mm]
\small
\textbf{Student's reflection:} \textit{``It becomes more necessary to fix specific parts of the code rather than defining the entire code.''}
\end{tcolorbox}

\subsection{RQ2: Usage of Multi-turn Conversations}

To investigate RQ2, we analyzed how multi-turn conversations supported students in solving problems of varying complexity. Specifically, we examined the total number of messages sent in their first successful conversation and the median character length of each message by its position (k-th message) within that conversation. This analysis provided insights into how students utilized multi-turn interactions to refine their prompts and achieve success.

Figure~\ref{fig.results_multiturn.success} highlights that students actively leveraged multi-turn conversations 
across all labs. In L7, most students succeeded within $2$ to $4$ messages, reflecting the simplicity of the problems. In L9, and more markedly in L10, as problem complexity increased, students made greater use of the multi-turn feature, with more than 25\% of the students who solved a problem in L10 requiring more than 5 messages to refine their solutions. This demonstrates that students did not rely on single, monolithic prompts to solve the problems but instead utilized smaller, iterative steps -- a more authentic and practical way to interact with generative AI tools \cite{DBLP:conf/chi/Vaithilingam0G22,denny2023conversing}.

Figure~\ref{fig.results_multiturn.effort} further shows that the first message was generally the longest, especially in L10, where problems required detailed framing and initial setup. Subsequent messages were shorter and more consistent in length, reflecting students' focused adjustments and iterative refinements as they progressed toward a solution. These patterns highlight the value of multi-turn conversations, enabling students to approach problem-solving incrementally and adapt their strategies based on model feedback, closely mirroring real-world interactions with AI systems.

\begin{tcolorbox}[colback=gray!10, colframe=black!40, rounded corners, boxsep=0mm, left=0.5mm, right=0.7mm, top=0.5mm,bottom=0.7mm]
\small
\textbf{Student's reflection:} \textit{``The final task took me multiple messages of correcting the code. Here, I found that if I wasn't specific enough then the AI would generate the wrong code.''}
\end{tcolorbox}

\subsection{RQ3: Usage of On-request Execution}

To evaluate the on-request code execution feature, we wish to understand how selective students are in deciding whether to execute the generated code.  Simply put, we analyzed whether students were more likely to execute code generated by the model when the code was correct compared to when the code was incorrect.  For every student prompt that resulted in code generation, we determined the correctness of that code and then analyzed whether or not the student chose to execute it.  

\begin{figure}[t!]
\centering
	\includegraphics[width=0.98\linewidth]{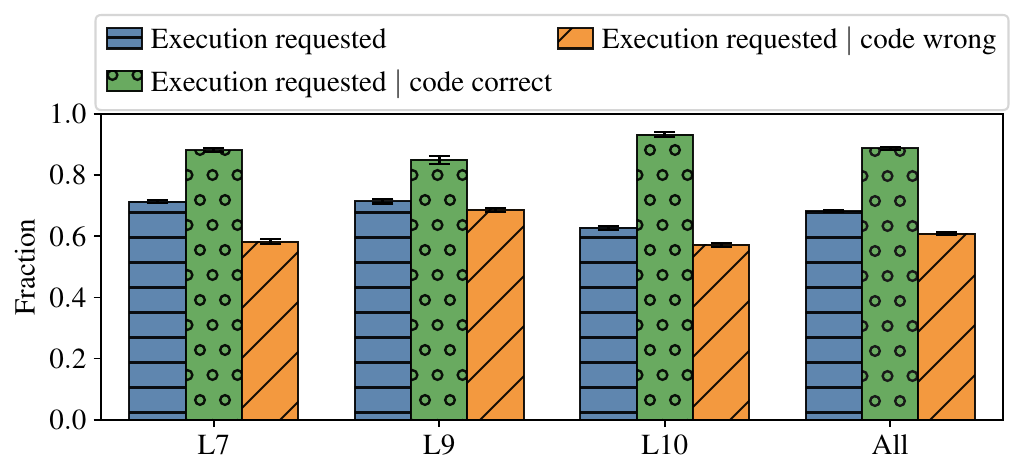}
    \vspace{-3mm}
     \caption{Fraction of code execution requests, including the overall requests, those with conditioning on code being correct, and those with conditioning on code being wrong.
     }
     \vspace{-2mm}
    \label{fig.results_codeexec}
\end{figure}

Figure~\ref{fig.results_codeexec} shows that across all messages that resulted in generated code, roughly $65$-$70\%$ of the time code execution was requested.  This provides some insight into how frequently students relied on running code to support their problem-solving process.
Moreover, students were much more likely to execute correct code than incorrect code, indicating that they applied judgment in assessing whether the generated code was ready for testing. This behavior demonstrates that on-request execution does encourage students to critically evaluate the output of the model before acting, fostering thoughtful and iterative problem-solving. Importantly, students consistently exhibited this judgment across all lab sessions.

\begin{tcolorbox}[colback=gray!10, colframe=black!40, rounded corners, boxsep=0mm, left=0.7mm, right=0.7mm, top=0.5mm,bottom=0.5mm]
\small
\textbf{Student's reflection:} \textit{``This prompt tool is good for understanding how to read code, which is something I need to work on. This task was very helpful.''}
\end{tcolorbox}

\subsection{Student Reflections}

To investigate student engagement with \emph{Prompt Programming}, we also gathered quantitative reflections. A common instructor concern relating to generative AI (GenAI) is the risk of student over-reliance \cite{prather2023navigating, sheard2024instructor, prather2024widening}. There may be a perception that platforms like ours, which support prompt programming, could reinforce this concern by leading students to believe that all problems can be solved through natural language prompting alone. This risk could be particularly pronounced if students were only exposed to trivial problems. However, we hypothesize that as problem complexity increases, it becomes more difficult to solve tasks solely through natural language prompting, even with multi-turn conversations.

\begin{figure}[t!]
    \centering
    \includegraphics[width=\linewidth,trim={16mm 7mm 23.3mm 9.5mm},clip]{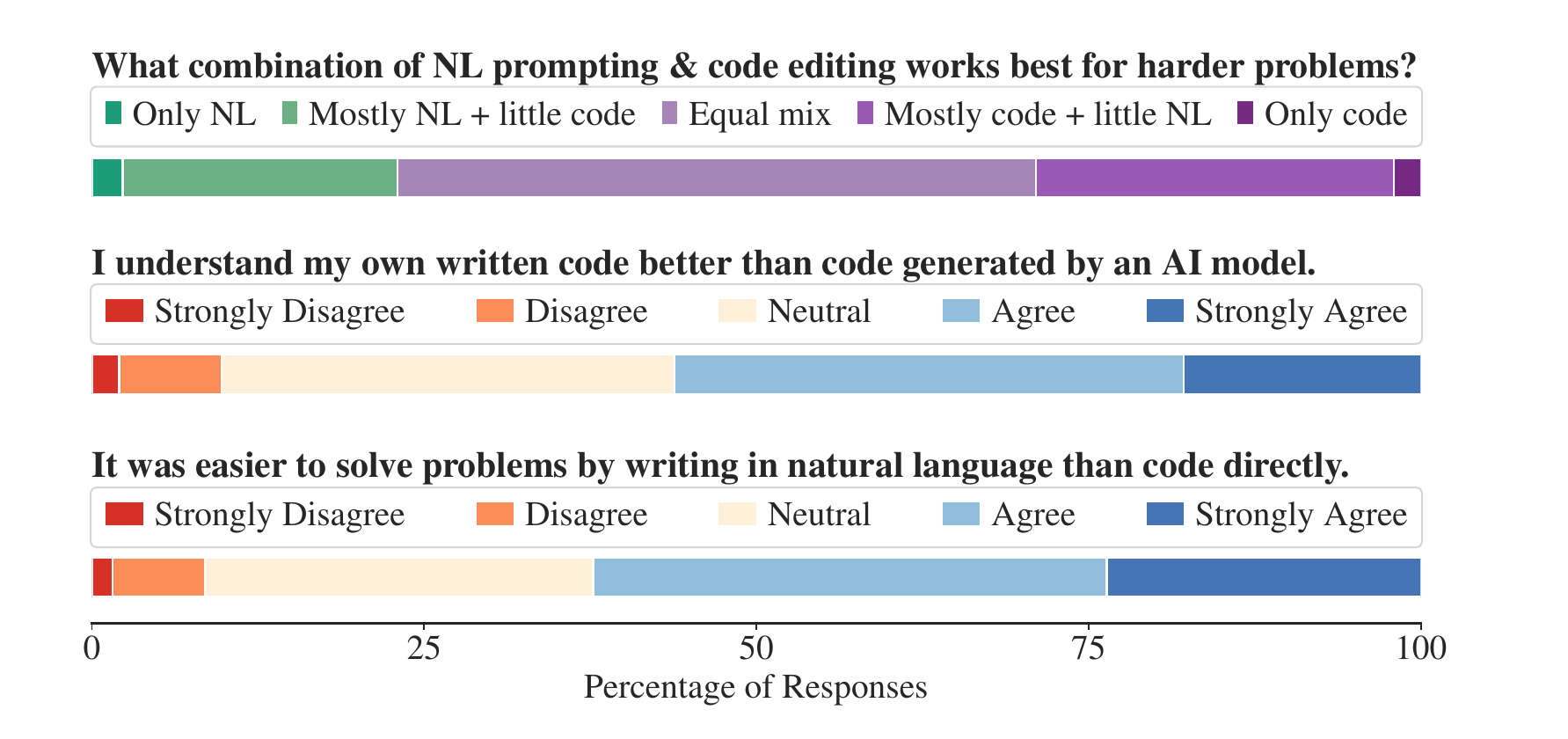}
    \vspace{-5mm}
    \caption{Preferences for combining natural language (NL) prompting and code editing, understanding AI-generated code, and problem-solving with NL prompting.}
\label{fig.results_reflections}
    \vspace{-3mm}
\end{figure}

\looseness-1To explore this, we asked students to indicate what they felt was the ideal combination of skills in natural language (NL)  prompting and manual code editing for solving harder problems. The results, shown in Figure~\ref{fig.results_reflections}, reveal a well-balanced response: the vast majority of students believed that an equal mix of these skills is most desirable. At the same time, students reported better understanding of their own code compared to AI-generated solutions, suggesting that working with AI-generated code requires additional effort to interpret and refine. Finally, students found NL prompting easier than direct coding, reflecting its ability to reduce the focus on syntax and low-level implementation details. These observations suggest teaching strategies should evolve to prioritize a hybrid approach, combining NL prompting and traditional programming skills.

\subsection{Limitations}
\label{sec:limitations}
\looseness-1Some limitations in our study suggest avenues for future work. First, the testing framework for \emph{multi-function} problems uses a single driver file that validates all functions together, which can lead to compilation errors if students run code before generating all required components. A more modular approach could enable independent validation of partial solutions. Second, prior exposure to L7 and L9 may have influenced student behavior in L10, such as restart frequency or message refinement. Third, we did not conduct an A/B study comparing single-turn and multi-turn conversations to isolate their impact on problem-solving. Finally, we did not examine how students' approaches and success relate to academic performance or how reliance on LLM-generated code affects students' long-term ability to independently generate and understand code. Each of these limitations highlights a direction for future investigation.

\section{Concluding Remarks}\label{sec.conclusion}

\looseness-1We introduced \emph{Prompt Programming}, a dialogue-based platform designed to help students develop the skill of crafting effective prompts while supporting iterative problem-solving with generative AI. Findings from its first use in a large introductory course demonstrate that students effectively used multi-turn conversations to refine solutions, particularly for complex multi-function problems. Students also exhibited judgment when using on-request code execution, running correct code more often than erroneous code. This highlights the platform's role in fostering skills in prompt writing, critical evaluation, and systematic debugging. Despite the focus on prompting, student reflections showed that they recognized the importance of integrating natural language prompting alongside more traditional coding skills when solving complex problems. 

\looseness-1\emph{Prompt Programming} is publicly available along with a curated set of programming exercises designed for educational use. The platform offers learners an accessible, interactive environment to practice prompt writing and iterative problem-solving in natural language. Future work will expand its capabilities by introducing direct code editing and broadening the variety of exercises. We also aim to explore multi-language interactions and measure the impact of \emph{Prompt Programming} on student learning outcomes. In the longer term, we aim to support educators to customize and assign exercises, and to enable learners to create and share their own problems, fostering a collaborative, community-driven learning environment.

\begin{tcolorbox}[colback=gray!10, colframe=black!40, rounded corners, boxsep=0mm, left=0.7mm, right=0.7mm, top=0.5mm,bottom=0.5mm]
\small
\textbf{Student's reflection:} \textit{``I really enjoyed this task!!! I find explaining how a code should run much easier than writing it for myself as I am dyslexic and don't have to worry about checking my variable names all match, etc.''}
\end{tcolorbox}
\begin{acks}
Funded/Cofunded by the European Union (ERC, TOPS, 101039090). Views and opinions expressed are however those of the author(s) only and do not necessarily reflect those of the European Union or the European Research Council. Neither the European Union nor the granting authority can be held responsible for them.
\end{acks}
\bibliographystyle{ACM-Reference-Format}
\bibliography{main}

\end{document}